\documentclass[12pt,aps,preprint]{revtex4}

\usepackage{amsmath}
\usepackage{mathrsfs}
\usepackage{bm}
\usepackage[pdftex]{hyperref}
\usepackage{graphicx}

\begin{document}
\begin{titlepage}	
\title{Elemental matrices for the finite element method in electromagnetics with quadratic triangular elements}
\author{E. Cojocaru}
\affiliation{Department of Theoretical Physics, 
Horia Hulubei National Institute of Physics and Nuclear Engineering, Magurele-Bucharest P.O.Box MG-6, 077125 Romania}

\email{ecojocaru@theory.nipne.ro}

\begin{abstract}
The finite element method has become a preeminent simulation technique in electromagnetics. For problems involving anisotropic media and metamaterials, proper algorithms should be developed. It has been proved that discretizing in quadratic triangular elements may lead to an improved accuracy. Here we present a collection of elemental matrices evaluated analytically for quadratic triangular elements. They could be useful for the finite element method in advanced electromagnetics. 
\end{abstract}

\maketitle
\end{titlepage}
\section{Introduction}
The finite element method is a numerical technique for obtaining approximate solutions to boundary-value  problems of mathematical physics. The method was developed and applied extensively for the analysis of electromagnetic problems \cite{1,2,3,4,5}.High-order vector finite elements have been developed which make it possible to obtain highly accurate and efficient solutions of vector wave equations \cite{2}. When applied to problems involving anisotropic media and metamaterials, since the most of available commercial packages cannot be applied, proper algorithms should be developed \cite{6}. An improved accuracy results with discretization in quadratic triangular elements \cite{2}. The evaluation of elemental matrices, although is a simple algebra, is rather cumbersome. Here we present a collection of elemental matrices evaluated analytically for quadratic triangular elements. 

\section{Basic relations}

\begin{figure}[ht]
\includegraphics[height=3cm]{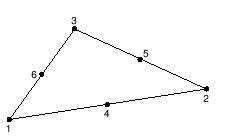}
\caption{\label{fig:fig_1}Quadratic triangular element.}
\end{figure}

We consider an optical medium with an arbitrary cross section $\Omega$ in the $xy$ plane. With a time dependence of the form $\textrm{exp}(j\omega t)$, where $\omega$ is the angular frequency, from Maxwell's equations the following vectorial wave function is derived
\begin{equation}
\label{eq:1}
\nabla \times([p]\nabla \times \phi) - k_0^2[q]\phi = 0
\end{equation}
where $k_0$ is the free-space wavenumber, $[p],[q]$ are related to the permittivity and permeability tensors, and $\phi$ denotes either the electric $\boldsymbol{E}$ or the magnetic $\boldsymbol{H}$ field. 
Dividing the cross section $\Omega$ into quadratic triangular elements, as shown in Fig.\ref{fig:fig_1}, we expand the transverse components $\phi_x,\phi_y$ and the axial components $\phi_z$ in each element as \cite{2}
\begin{equation}
\label{eq:2}
\phi=\begin{bmatrix}\phi_x\\ \phi_y\\ \phi_z \end{bmatrix} = \begin{bmatrix} [U]^T[\phi_t]_e\\
[V]^T[\phi_t]_e\\j[N]^T[\phi_z]_e \end{bmatrix}
\end{equation}
where $[\phi_t]_e$ is the transverse tangential field, $[\phi_z]_e$ is the longitudinal nodal field of each element, $[U],[V],[N]$ are shape function vectors, and $T$ denotes a transpose. The shape function $[N]$ has six components $[N]=[N_1 N_2 N_3 N_4 N_5 N_6]^T$ which are expressed
in terms of area coordinates $L_1, L_2, L_3$,
\begin{eqnarray}
\label{eq:3}
N_1&\!\!=\!\!&L_1(2L_1 - 1) \nonumber \\
N_2&\!\!=\!\!&L_2(2L_2 - 1) \nonumber \\
N_3&\!\!=\!\!&L_3(2L_3 - 1) \nonumber \\
N_4&\!\!=\!\!&4L_1 L_2 \nonumber \\
N_5&\!\!=\!\!&4L_2 L_3 \nonumber \\
N_6&\!\!=\!\!&4L_3 L_1 
\end{eqnarray}
The area coordinates $L_k(k = 1,2,3)$ are given by
\begin{equation}
\label{eq:4}
\begin{bmatrix}L_1 \\ L_2 \\ L_3
\end{bmatrix} = \frac{1}{2A_e}\begin{bmatrix}a_1 & b_1 & c_1 \\ a_2 & b_2 & c_2 \\a_3 & b_3 & c_3
\end{bmatrix}
\end{equation}
The area of the element $A_e$ is given by
\begin{equation}
\label{eq:5}
2A_e=\begin{vmatrix} 1 & 1 & 1\\ x_1 & x_2 & x_3 \\ y_1 & y_2 & y_3 
\end{vmatrix}
\end{equation}
where $x_k,~y_k$ are the Cartesian coordinates of the corner points 1 to 3 of the triangle, and the subscripts $k,l,m$ progress cyclically around the three corners. The coefficients $a_k,b_k,c_k$ are given by
\begin{eqnarray}
\label{eq:6}
a_k&\!\!=\!\!&x_l y_m - x_m y_l \nonumber \\
b_k&\!\!=\!\!&y_l -y_m \nonumber \\
c_k&\!\!=\!\!&x_m -x_l
\end{eqnarray}
The shape functions $[U]$ and $[V]$ are given by \cite{2}
\begin{equation}
\label{eq:7}
[U] = \frac{1}{2A_e}\begin{bmatrix}l_1 b_2 L_1 \\ l_2 b_3 L_2 \\ l_3 b_1 L_3 \\
 -l_1 b_1 L_2 \\ -l_2 b_2 L_3 \\ -l_3 b_3 L_1
 \end{bmatrix} \qquad  [V] = \frac{1}{2A_e}\begin{bmatrix}l_1 c_2 L_1 \\ l_2 c_3 L_2 \\ l_3 c_1 L_3 \\
 -l_1 c_1 L_2 \\ -l_2 c_2 L_3 \\ -l_3 c_3 L_1
 \end{bmatrix}
 \end{equation}
with $|l_k| = \sqrt{b_m^2+c_m^2}$ where the subscripts $k,l,m$ progress cyclically around the three corners of the triangle, but $l_k$ includes a sign which is defined differently in \cite{2,5,7}.
When vector shape functions such as those given above are employed to represent a vector field in a finite element solution of vector wave equation, it is necessary to consider evaluation of the associated elemental matrices. These integrals can be evaluated analytically for triangular elements. In case of simple triangular elements, analytical relations for elemental matrices can be found in literature \cite{1}. For quadratic triangular elements the integration is generally more involved. Analytical relations presented in \cite{2} are rather complex. Different forms of elemental matrices could be useful in order to check their correctness when they are implemented in a developed algorithm. In the following we present a collection of elemental matrices for quadratic triangular elements.

\section{Elemental matrices evaluated for quadratic triangular elements}

We used the following integration formula for the area coordinates \cite{3}
\begin{equation}
\label{8}
\iint_e L_1^i L_2^j L_3^k \mathrm{d} x \mathrm{d} y =\frac{i!j!k!}{(i+j+k+2)!}2A_e \qquad (i,j,k=0,1,2,3,\dots)
\end{equation}
Sometimes, relations becomes simpler if we take into account that
\begin{equation}
\label{eq:9}
b_1+b_2+b_3=0, \qquad c_1+c_2+c_3=0
\end{equation}
We begin with the most simple elemental matrix, $\iint_e [N][N]^T \mathrm{d} x \mathrm{d} y$, which can be found also in \cite{2,3,4}
\begin{equation}
\label{eq:10}
\iint_e [N][N]^T \mathrm{d} x \mathrm{d} y =\frac{A_e}{180}\begin{bmatrix}6 & -1 & -1 &  0 & -4 &  0 \\
                                 -1 &  6 & -1 &  0 &  0 & -4 \\
                                 -1 & -1 &  6 & -4 &  0 &  0 \\
                                  0 &  0 & -4 & 32 & 16 & 16 \\
                                 -4 &  0 &  0 & 16 & 32 & 16 \\
                                  0 & -4 &  0 & 16 & 16 & 32 
\end{bmatrix}
\end{equation}
\begin{eqnarray}
\label{eq:11}
&&\iint_e \frac{\partial[N]}{\partial x}\frac{\partial[N]^T}{\partial x}\mathrm{d} x \mathrm{d} y   
 = \frac{1}{3A_e} \times \nonumber \\
&&\!\!\begin{bmatrix}
\frac{3b_1^2}{4}    &  \frac{-b_1 b_2}{4} &   \frac{-b_1 b_3}{4} &  b_1 b_2   &        0         &   b_1 b_3 \\
\frac{-b_1 b_2}{4}  &  \frac{3b_2^2}{4}   &   \frac{-b_2 b_3}{4} &  b_1 b_2   &        b_2 b_3  &   0        \\
\frac{-b_1 b_3}{4}  &  \frac{-b_2 b_3}{4} &   \frac{3b_3^2}{4}   &  0          &        b_2 b_3  &   b_1 b_3 \\
b_1 b_2  &  b_1 b_2 &   0        &  2(b_1^2+b_2^2+b_1 b_2)  &         2b_1 b_3 & 2b_2 b_3 \\
0         &  b_2 b_3 &   b_2 b_3 &  2b_1 b_3 & 2(b_2^2+b_3^2+b_2 b_3) & 2b_1 b_2 \\
b_1 b_3     &   0   &            b_1 b_3 & 2b_2 b_3 & 2b_1 b_2 & 2(b_1^2+b_3^2+b_1 b_3)
\end{bmatrix}
\end{eqnarray}
\begin{eqnarray}
\label{eq:12}
&&\iint_e \frac{\partial[N]}{\partial y}\frac{\partial[N]^T}{\partial y}\mathrm{d} x \mathrm{d} y = \frac{1}{3A_e} \times \nonumber \\
&&\begin{bmatrix}
\frac{3c_1^2}{4}   &     \frac{-c_1 c_2}{4}    &   \frac{-c_1 c_3}{4}   &   c_1 c_2  &      0          &     c_1 c_3  \\
\frac{-c_1 c_2}{4} &     \frac{3c_2^2}{4}      &   \frac{-c_2 c_3}{4}   &   c_1 c_2  &      c_2 c_3   &     0        \\
\frac{-c_1 c_3}{4} &  \frac{-c_2 c_3}{4} & \frac{3c_3^2}{4}  &   0    &   c_2 c_3   &     c_1 c_3 \\
c_1 c_2 &     c_1 c_2    &   0    &   2(c_1^2+c_2^2+c_1 c_2)  & 2c_1 c_3  &   2c_2 c_3 \\
0   &   c_2 c_3 &   c_2 c_3   &   2c_1 c_3 & 2(c_2^2+c_3^2+c_2 c_3)  &   2c_1 c_2 \\
c_1 c_3 &     0   &   c_1 c_3   &   2c_2 c_3 & 2c_1 c_2   & 2(c_1^2+c_3^2+c_1 c_3)
\end{bmatrix}
\end{eqnarray}
Note that Eqs.(\ref{eq:11}) and (\ref{eq:12}) are given also in \cite{2,3,4}, but here they are simplified by using Eq.(\ref{eq:9}).

\begin{eqnarray}
\label{eq:13}
&&\iint_e \frac{\partial[N]}{\partial y}\frac{\partial[N]^T}{\partial x}\mathrm{d} x \mathrm{d} y = \frac{1}{3A_e} \times \nonumber \\
&&\begin{bmatrix}
\frac{3b_1 c_1}{4}  &    \frac{-b_2 c_1}{4}   &    \frac{-b_3 c_1}{4}   &    b_2 c_1   &    0   &   b_3 c_1 \\
\frac{-b_1 c_2}{4}  &    \frac{3b_2 c_2}{4}   & \frac{-b_3 c_2}{4}   &  b_1 c_2   &  b_3 c_2   &   0 \\
\frac{-b_1 c_3}{4}  & \frac{-b_2 c_3}{4}   & \frac{3b_3 c_3}{4}   &    0  &  b_2 c_3   &   b_1 c_3 \\
b_1 c_2 & b_2 c_1 & 0 & b_1 c_1+b_2 c_2+b_3 c_3 & b_3 c_1+b_1 c_3 & b_3 c_2+b_2 c_3 \\
0 & b_2 c_3 & b_3 c_2 & b_1 c_3+b_3 c_1 &b_1 c_1+b_2 c_2+b_3 c_3& b_1 c_2+b_2 c_1 \\
b_1 c_3  &  0  &  b_3 c_1 & b_2 c_3+b_3 c_2 & b_2 c_1+b_1 c_2 & b_1 c_1+b_2 c_2+b_3 c_3
\end{bmatrix}
\end{eqnarray}
\begin{equation}
\label{eq:14}
\iint_e [N]\frac{\partial[N]^T}{\partial x}\mathrm{d} x \mathrm{d} y = \frac{1}{30}\begin{bmatrix}
    2b_1  &  -b_2  &  -b_3    &  -b_1+2b_2    &  -b_2-b_3     &  -b_1+2b_3   \\
    -b_1  &  2b_2  &  -b_3    &  -b_2+2b_1    &  -b_2+2b_3    &  -b_1-b_3    \\
    -b_1  &  -b_2  &  2b_3    &  -b_1-b_2     &  -b_3+2b_2    &  -b_3+2b_1   \\
    3b_1  &  3b_2  &  -b_3    &  8(b_1+b_2)   &  4(b_2+2b_3)  &  4(b_1+2b_3) \\
    -b_1  &  3b_2  &  3b_3    &  4(b_2+2b_1)  &  8(b_2+b_3)   &  4(b_3+2b_1) \\
    3b_1  &  -b_2  &  3b_3    &  4(b_1+2b_2)  &  4(b_3+2b_2)  &  8(b_1+b_3)
\end{bmatrix}
\end{equation}       
\begin{equation}
\label{eq:15}
\iint_e [N]\frac{\partial[N]^T}{\partial y}\mathrm{d} x \mathrm{d} y = \frac{1}{30}\begin{bmatrix}
2c_1   &  -c_2   &  -c_3   &   -c_1+2c_2    &  -c_2-c_3    &   -c_1+2c_3    \\
-c_1   &  2c_2   &  -c_3   &   -c_2+2c_1    &  -c_2+2c_3   &   -c_1-c_3     \\
-c_1   &  -c_2   &  2c_3   &   -c_1-c_2     &  -c_3+2c_2   &   -c_3+2c_1    \\
3c_1   &  3c_2   &  -c_3   &   8(c_1+c_2)   &  4(c_2+2c_3) &   4(c_1+2c_3)  \\
-c_1   &  3c_2   &  3c_3   &   4(c_2+2c_1)  &  8(c_2+c_3)  &   4(c_3+2c_1)  \\
3c_1   &  -c_2   &  3c_3   &   4(c_1+2c_2)  &  4(c_3+2c_2) &   8(c_1+c_3)   
\end{bmatrix}
\end{equation}      
\begin{eqnarray}
\label{eq:16}
&&\iint_e [U]\frac{\partial[N]^T}{\partial x}\mathrm{d} x \mathrm{d} y = \frac{1}{12A_e} \times \nonumber \\
&&\begin{bmatrix}
l_1 b_1 b_2 &   0     &   0   &  l_1 b_2(b_1+2b_2)  &  l_1 b_2(b_2+b_3)   &  l_1 b_2(b_1+2b_3)  \\
0   &   l_2 b_2 b_3   &   0   &  l_2 b_3(b_2+2b_1)  &  l_2 b_3(b_2+2b_3)  &  l_2 b_3(b_1+b_3)   \\ 
0   &   0   &   l_3 b_1 b_3   &  l_3 b_1(b_1+b_2)   &  l_3 b_1(b_3+2b_2)  &  l_3 b_1(b_3+2b_1)  \\
0   &   -l_1 b_1 b_2  &   0   &  -l_1 b_1(b_2+2b_1) &  -l_1 b_1(b_2+2b_3) &  -l_1 b_1(b_1+b_3)  \\
0   &   0   &   -l_2 b_2 b_3  &  -l_2 b_2(b_1+b_2)  &  -l_2 b_2(b_3+2b_2) &  -l_2 b_2(b_3+2b_1) \\
-l_3 b_1 b_3  & 0     &   0   &  -l_3 b_3(b_1+2b_2) &  -l_3 b_3(b_2+b_3)  &  -l_3 b_3(b_1+2b_3)
\end{bmatrix}
\end{eqnarray} 
\begin{eqnarray}
\label{eq:17}
&&\iint_e [V]\frac{\partial[N]^T}{\partial x}\mathrm{d} x \mathrm{d} y = \frac{1}{12A_e} \times \nonumber \\
&&\begin{bmatrix}
l_1 b_1 c_2   &   0    &    0   &  l_1 c_2(b_1+2b_2)  &    l_1 c_2(b_2+b_3)  &   l_1 c_2(b_1+2b_3) \\
0  &   l_2 b_2 c_3     &    0   &  l_2 c_3(b_2+2b_1)  &    l_2 c_3(b_2+2b_3) &   l_2 c_3(b_1+b_3) \\
0  &   0      &   l_3 b_3 c_1   &  l_3 c_1(b_1+b_2)   &    l_3 c_1(b_3+2b_2) &   l_3 c_1(b_3+2b_1)\\
0  &   -l_1 b_2 c_1    &    0   &  -l_1 c_1(b_2+2b_1) &    -l_1 c_1(b_2+2b_3)&   -l_1 c_1(b_1+b_3)\\ 
0  &   0      &   -l_2 b_3 c_2  &  -l_2 c_2(b_1+b_2)  &    -l_2 c_2(b_3+2b_2)&   -l_2 c_2(b_3+2b_1)\\
-l_3 b_1 c_3  &   0    &    0   &  -l_3 c_3(b_1+2b_2) &    -l_3 c_3(b_2+b_3) &   -l_3 c_3(b_1+2b_3)
\end{bmatrix}
\end{eqnarray} 
\begin{eqnarray}
\label{eq:18}
&&\iint_e [U]\frac{\partial[N]^T}{\partial y}\mathrm{d} x \mathrm{d} y = \frac{1}{12A_e} 
\times \nonumber \\
&&\begin{bmatrix}
l_1 b_2 c_1  &   0   &   0  &  l_1 b_2(c_1+2c_2)  &  l_1 b_2(c_2+c_3)  &  l_1 b_2(c_1+2c_3)  \\
0  &  l_2 b_3 c_2    &   0  &  l_2 b_3(c_2+2c_1)  &  l_2 b_3(c_2+2c_3) &  l_2 b_3(c_1+c_3)   \\
0  &  0   &  l_3 b_1 c_3    &  l_3 b_1(c_1+c_2)   &  l_3 b_1(c_3+2c_2) &  l_3 b_1(c_3+2c_1)  \\
0  &  -l_1 b_1 c_2   &   0  &  -l_1 b_1(c_2+2c_1) &  -l_1 b_1(c_2+2c_3)&  -l_1 b_1(c_1+c_3)  \\
0  &  0   &  -l_2 b_2 c_3   &  -l_2 b_2(c_1+c_2)  &  -l_2 b_2(c_3+2c_2)&  -l_2 b_2(c_3+2c_1) \\
-l_3 b_3 c_1  &   0  &   0  &  -l_3 b_3(c_1+2c_2) &  -l_3 b_3(c_2+c_3) &  -l_3 b_3(c_1+2c_3)
\end{bmatrix}
\end{eqnarray} 
\begin{eqnarray}
\label{eq:19}
&&\iint_e [V]\frac{\partial[N]^T}{\partial y}\mathrm{d} x \mathrm{d} y = \frac{1}{12A_e}
\times \nonumber \\
&&\begin{bmatrix}
l_1 c_1 c_2   &   0      &   0  &  l_1 c_2(c_1+2c_2)  &  l_1 c_2(c_2+c_3)    &  l_1 c_2(c_1+2c_3)    \\
0   &     l_2 c_2 c_3    &   0  &  l_2 c_3(c_2+2c_1)  &  l_2 c_3(c_2+2c_3)   &  l_2 c_3(c_1+c_3)     \\
0   &     0   &   l_3 c_1 c_3   &  l_3 c_1(c_1+c_2)   &  l_3 c_1(c_3+2c_2)   &  l_3 c_1(c_3+2c_1)    \\   
0   &     -l_1 c_1 c_2   &      0  &  -l_1 c_1(c_2+2c_1) &  -l_1 c_1(c_2+2c_3)  &  -l_1 c_1(c_1+c_3) \\
0   &     0   &   -l_2 c_2 c_3  &  -l_2 c_2(c_1+c_2)  &  -l_2 c_2(c_3+2c_2)  &  -l_2 c_2(c_3+2c_1)   \\
-l_3 c_1 c_3  &   0   &      0  &  -l_3 c_3(c_1+2c_2) &  -l_3 c_3(c_2+c_3)   &  -l_3 c_3(c_1+2c_3)
\end{bmatrix}
\end{eqnarray} 
\begin{eqnarray}
\label{eq:20}
&&\iint_e [U][U]^T \mathrm{d} x \mathrm{d} y = \frac{1}{48A_e} \times \nonumber \\
&&\begin{bmatrix}
2l_1^2 b_2^2  & l_1 l_2 b_2 b_3 & l_1 l_3 b_1 b_2 &  -l_1^2 b_1 b_2 &  -l_1 l_2 b_2^2 &  -2l_1 l_3 b_2 b_3 \\
l_1 l_2 b_2 b_3 & 2l_2^2 b_3^2  & l_2 l_3 b_1 b_3 &  -2l_1 l_2 b_1 b_3 &  -l_2^2 b_2 b_3 & -l_2 l_3 b_3^2 \\
l_1 l_3 b_1 b_2 & l_2 l_3 b_1 b_3 & 2l_3^2 b_1^2  & -l_1 l_3 b_1^2 & -2l_2 l_3 b_1 b_2 &  -l_3^2 b_1 b_3 \\
-l_1^2 b_1 b_2  & -2l_1 l_2 b_1 b_3 &  -l_1 l_3 b_1^2 & 2l_1^2 b_1^2 &  l_1 l_2 b_1 b_2 & l_1 l_3 b_1 b_3\\
-l_1 l_2 b_2^2 & -l_2^2 b_2 b_3 & -2l_2 l_3 b_1 b_2 & l_1 l_2 b_1 b_2 & 2l_2^2 b_2^2 &  l_2 l_3 b_2 b_3 \\
-2l_1 l_3 b_2 b_3 &  -l_2 l_3 b_3^2 & -l_3^2 b_1 b_3 & l_1 l_3 b_1 b_3 & l_2 l_3 b_2 b_3 & 2l_3^2 b_3^2
\end{bmatrix}
\end{eqnarray}
\begin{eqnarray}
\label{eq:21}
&&\iint_e [V][V]^T \mathrm{d} x \mathrm{d} y = \frac{1}{48A_e} \times \nonumber \\
&&\begin{bmatrix}     
2l_1^2 c_2^2 & l_1 l_2 c_2 c_3 &  l_1 l_3 c_1 c_2 & -l_1^2 c_1 c_2 & -l_1 l_2 c_2^2  &   -2l_1 l_3 c_2 c_3 \\
l_1 l_2 c_2 c_3 & 2l_2^2 c_3^2 &  l_2 l_3 c_1 c_3 & -2l_1 l_2 c_1 c_3 &  -l_2^2 c_2 c_3 &   -l_2 l_3 c_3^2 \\
l_1 l_3 c_1 c_2 & l_2 l_3 c_1 c_3 &   2l_3^2 c_1^2 & -l_1 l_3 c_1^2 &  -2l_2 l_3 c_1 c_2 & -l_3^2 c_1 c_3 \\
-l_1^2 c_1 c_2 & -2l_1 l_2 c_1 c_3 &  -l_1 l_3 c_1^2 & 2l_1^2 c_1^2 &  l_1 l_2 c_1 c_2 &  l_1 l_3 c_1 c_3 \\
-l_1 l_2 c_2^2 & -l_2^2 c_2 c_3 &    -2l_2 l_3 c_1 c_2 & l_1 l_2 c_1 c_2 &  2l_2^2 c_2^2 &  l_2 l_3 c_2 c_3\\
-2l_1 l_3 c_2 c_3 & -l_2 l_3 c_3^2  & -l_3^2 c_1 c_3 & l_1 l_3 c_1 c_3 &  l_2 l_3 c_2 c_3 &  2l_3^2 c_3^2
\end{bmatrix}
\end{eqnarray}
\begin{eqnarray}
\label{eq:22}
&&\iint_e [U][V]^T \mathrm{d} x \mathrm{d} y = \frac{1}{48A_e} \times \nonumber \\
&&\begin{bmatrix}  
2l_1^2 b_2 c_2 &  l_1 l_2 b_2 c_3 &  l_1 l_3 b_2 c_1 &-l_1^2 b_2 c_1 &  -l_1 l_2 b_2 c_2 &  -2l_1 l_3 b_2 c_3\\
l_1 l_2 b_3 c_2 & 2l_2^2 b_3 c_3 &   l_2 l_3 b_3 c_1&-2l_1 l_2 b_3 c_1 & -l_2^2 b_3 c_2& -l_2 l_3 b_3 c_3\\
l_1 l_3 b_1 c_2& l_2 l_3 b_1 c_3& 2l_3^2 b_1 c_1& -l_1 l_3 b_1 c_1&  -2l_2 l_3 b_1 c_2& -l_3^2 b_1 c_3\\
-l_1^2 b_1 c_2&  -2l_1 l_2 b_1 c_3&-l_1 l_3 b_1 c_1& 2l_1^2 b_1 c_1& l_1 l_2 b_1 c_2&  l_1 l_3 b_1 c_3\\
-l_1 l_2 b_2 c_2&  -l_2^2 b_2 c_3& -2l_2 l_3 b_2 c_1& l_1 l_2 b_2 c_1&   2l_2^2 b_2 c_2& l_2 l_3 b_2 c_3\\
-2l_1 l_3 b_3 c_2 & -l_2 l_3 b_3 c_3& -l_3^2 b_3 c_1 &l_1 l_3 b_3 c_1&   l_2 l_3 b_3 c_2&     2l_3^2 b_3 c_3
\end{bmatrix}
\end{eqnarray}   
\begin{equation}
\label{eq:23}    
\iint_e \frac{\partial[U]}{\partial y}\frac{\partial[U]^T}{\partial y}\mathrm{d} x \mathrm{d} y = \frac{1}{16A_e^3}\begin{bmatrix} a_{11} & a_{12}  & a_{13} & a_{14} & a_{15} & a_{16} \\
    a_{12} & a_{22}  & a_{23} & a_{24} & a_{25} & a_{26} \\
    a_{13} & a_{23}  & a_{33} & a_{34} & a_{35} & a_{36} \\
    a_{14} & a_{24}  & a_{34} & a_{44} & a_{45} & a_{46} \\
    a_{15} & a_{25}  & a_{35} & a_{45} & a_{55} & a_{56} \\
    a_{16} & a_{26}  & a_{36} & a_{46} & a_{56} & a_{66} 
\end{bmatrix}
\end{equation} 
\begin{eqnarray}
\label{eq:24}    
a_{11}&\!\!=\!\!&l_1^2 b_2^2 c_1^2, \qquad  a_{12} = l_1 l_2 b_2 b_3 c_1 c_2, \qquad  a_{13} = l_1 l_3 b_1 b_2 c_1 c_3,  \nonumber \\
a_{14}&\!\!=\!\!&-l_1^2 b_1 b_2 c_1 c_2, \qquad a_{15} = -l_1 l_2 b_2^2 c_1 c_3, \qquad  a_{16} = -l_1 l_3 b_2 b_3 c_1^2,  \nonumber \\
a_{22}&\!\!=\!\!&l_2^2 b_3^2 c_2^2, \qquad    a_{23} = l_2 l_3 b_1 b_3 c_2 c_3, \qquad a_{24} =
-l_1 l_2 b_1 b_3 c_2^2, \nonumber  \\
a_{25}&\!\!=\!\!&-l_2^2 b_2 b_3 c_2 c_3, \qquad  a_{26} = -l_2 l_3 b_3^2 c_1 c_2, \qquad a_{33} =
l_3^2 b_1^2 c_3^2,  \nonumber \\
a_{34}&\!\!=\!\!&-l_1 l_3 b_1^2 c_2 c_3, \qquad  a_{35} = -l_2 l_3 b_1 b_2 c_3^2, \qquad a_{36} = -l_3^2 b_1 
b_3 c_1 c_3,  \nonumber \\
a_{44}&\!\!=\!\!&l_1^2 b_1^2 c_2^2, \qquad    a_{45} = l_1 l_ 2 b_1 b_2 c_2 c_3, \qquad a_{46} = l_1 l_3 b_1 b_3 c_1 c_2,  \nonumber \\
a_{55}&\!\!=\!\!&l_2^2 b_2^2 c_3^2, \qquad  a_{56} = l_2 l_3 b_2 b_3 c_1 c_3, \qquad  a_{66} = l_3^2 b_3^2 c_1^2.
\end{eqnarray}
The matrix $\iint_e \frac{\partial[V]}{\partial x}\frac{\partial[V]^T}{\partial x}\mathrm{d} x \mathrm{d} y$ is expressed like in Eqs.(\ref{eq:23}) and (\ref{eq:24}) by applying the transformation $b\!\leftrightarrow\!c$. One obtains 
\begin{eqnarray}
\label{eq:25}    
a_{11}&\!\!=\!\!&l_1^2 c_2^2 b_1^2, \qquad  a_{12} = l_1 l_2 c_2 c_3 b_1 b_2, \qquad  a_{13} = l_1 l_3 c_1 c_2 b_1 b_3,  \nonumber \\
a_{14}&\!\!=\!\!&-l_1^2 c_1 c_2 b_1 b_2, \qquad a_{15} = -l_1 l_2 c_2^2 b_1 b_3, \qquad  a_{16} = -l_1 l_3 c_2 c_3 b_1^2,  \nonumber \\
a_{22}&\!\!=\!\!&l_2^2 c_3^2 b_2^2, \qquad    a_{23} = l_2 l_3 c_1 c_3 b_2 b_3, \qquad a_{24} =
-l_1 l_2 c_1 c_3 b_2^2, \nonumber  \\
a_{25}&\!\!=\!\!&-l_2^2 c_2 c_3 b_2 b_3, \qquad  a_{26} = -l_2 l_3 c_3^2 b_1 b_2, \qquad a_{33} =
l_3^2 b_1^2 c_3^2,  \nonumber \\
a_{34}&\!\!=\!\!&-l_1 l_3 c_1^2 b_2 b_3, \qquad  a_{35} = -l_2 l_3 c_1 c_2 b_3^2, \qquad a_{36} = -l_3^2 c_1 
c_3 b_1 b_3,  \nonumber \\
a_{44}&\!\!=\!\!&l_1^2 c_1^2 b_2^2, \qquad    a_{45} = l_1 l_ 2 c_1 c_2 b_2 b_3, \qquad a_{46} = l_1 l_3 c_1 c_3 b_1 b_2,  \nonumber \\
a_{55}&\!\!=\!\!&l_2^2 c_2^2 b_3^2, \qquad  a_{56} = l_2 l_3 c_2 c_3 b_1 b_3, \qquad  a_{66} = l_3^2 c_3^2 b_1^2.
\end{eqnarray}     
\begin{equation}
\label{eq:26}
\iint_e \frac{\partial[U]}{\partial y}\frac{\partial[V]^T}{\partial x}\mathrm{d} x \mathrm{d} y = \frac{1}{16A_e^3}\begin{bmatrix} a_{11} & a_{12}  & a_{13} & a_{14} & a_{15} & a_{16} \\
    a_{21} & a_{22}  & a_{23} & a_{24} & a_{25} & a_{26} \\
    a_{31} & a_{32}  & a_{33} & a_{34} & a_{35} & a_{36} \\
    a_{41} & a_{42}  & a_{43} & a_{44} & a_{45} & a_{46} \\
    a_{51} & a_{52}  & a_{53} & a_{54} & a_{55} & a_{56} \\
    a_{61} & a_{62}  & a_{63} & a_{64} & a_{65} & a_{66} 
\end{bmatrix}
\end{equation} 
\begin{eqnarray}
\label{eq:27}    
a_{11}&\!\!=\!\!&l_1^2 b_1 b_2 c_1 c_2,   \qquad   a_{12} = l_1 l_2 b_2^2 c_1 c_3, \qquad   a_{13} = l_1 l_3 b_2 b_3 c_1^2, \nonumber \\
a_{14}&\!\!=\!\!&-l_1^2 b_2^2 c_1^2,  \qquad     a_{15} = -l_1 l_2 b_2 b_3 c_1 c_2, \qquad  a_{16} = -l_1 l_3 b_1 b_2 c_1 c_3,  \nonumber \\
a_{21}&\!\!=\!\!& l_1 l_2 b_1 b_3 c_2^2,  \qquad    a_{22} = l_2^2 b_2 b_3 c_2 c_3,  \qquad   a_{23} = l_2 l_3 b_3^2 c_1 c_2, \nonumber \\
a_{24}&\!\!=\!\!&-l_1 l_2 b_2 b_3 c_1 c_2, \qquad    a_{25} = -l_2^2 b_3^2 c_2^2,  \qquad     a_{26} = -l_2 l_3 b_1 b_3 c_2 c_3, \nonumber \\
a_{31}&\!\!=\!\!& l_1 l_3 b_1^2 c_2 c_3,  \qquad     a_{32} = l_2 l_3 b_1 b_2 c_3^2,  \qquad    a_{33} = l_3^2 b_1 b_3 c_1 c_3,  \nonumber \\
a_{34}&\!\!=\!\!&-l_1 l_3 b_1 b_2 c_1 c_3,  \qquad   a_{35} = -l_2 l_3 b_1 b_3 c_2 c_3, \qquad   a_{36} = -l_3^2 b_1^2 c_3^2, \nonumber \\
a_{41}&\!\!=\!\!&-l_1^2 b_1^2 c_2^2,   \qquad     a_{42} = -l_1 l_2 b_1 b_2 c_2 c_3, \qquad   a_{43} = -l_1 l_3 b_1 b_3 c_1 c_2, \nonumber \\
a_{44}&\!\!=\!\!& l_1^2 b_1 b_2 c_1 c_2,  \qquad     a_{45} = l_1 l_2 b_1 b_3 c_2^2,  \qquad    a_{46} = l_1 l_3 b_1^2 c_2 c_3, \nonumber \\
a_{51}&\!\!=\!\!&-l_1 l_2 b_1 b_2 c_2 c_3, \qquad    a_{52} = -l_2^2 b_2^2 c_3^2,    \qquad   a_{53} = -l_2 l_3 b_2 b_3 c_1 c_3, \nonumber \\
a_{54}&\!\!=\!\!& l_1 l_2 b_2^2 c_1 c_3,  \qquad     a_{55} = l_2^2 b_2 b_3 c_2 c_3,  \qquad    a_{56} = l_2 l_3 b_1 b_2 c_3^2, \nonumber \\
a_{61}&\!\!=\!\!&-l_1 l_3 b_1 b_3 c_1 c_2,  \qquad   a_{62} = -l_2 l_3 b_2 b_3 c_1 c_3, \qquad   a_{63} = -l_3^2 b_3^2 c_1^2, \nonumber \\
a_{64}&\!\!=\!\!& l_1 l_3 b_2 b_3 c_1^2,  \qquad     a_{65} = l_2 l_3 b_3^2 c_1 c_2,  \qquad    a_{66} = l_3^2 b_1 b_3 c_1 c_3.
\end{eqnarray}


\begin{thebibliography}{7}

\bibitem{1}
M.~Koshiba and K.~Inoue, ``Simple and efficient finite-element analysis of microwave and optical waveguides,'' IEEE Trans. Microwave Theory Tech., vol. 40, no. 2, pp. 371--377, 1992.

\bibitem{2}
M.~Koshiba, S.~Maruyama, and K. Hirayama, ``A vector finite element method with the high-order mixed-interpolation-type triangular elements for optical waveguiding problems,'' J. Lightw. Technol., vol.12, no. 3, pp. 495--502, 1994.

\bibitem{3}
K.~Kawano and T.~Kitoh, \textit{Introduction to optical waveguide analysis. Solving Maxwell's equations and the Schrodinger equation}, John Wiley \& Sons, New York 2001.

\bibitem{4}
J.~Jin, \textit{The finite element method in electromagnetics}, Second Edition, John Wiley \& Sons, New York 2002.

\bibitem{5}
H.~Deng, \textit{Design and characterization of silicon-on-insulator passive polarization converter with finite-element analysis}, Ph.~D. thesis, University of Waterloo, Ontario, Canada, 2005.

\bibitem{6}
Y.~W.~Kwon and H.~Bang, \textit{The finite element method using Matlab}, CRC Press, New York 2000.

\bibitem{7}
D.~R.~Tanner and A.~F.~Peterson, ``Vector expansion functions for the numerical solution of Maxwell's equations,'' Microwave and Optical Technology Letters, vol. 2, no. 9, pp. 331--334, 1989.

\end{thebibliography}
\end{document}